\documentclass[aip,
preprint,%
]{revtex4-1}

\usepackage{lmodern}
\usepackage{graphicx}% Include figure files
\usepackage{dcolumn}% Align table columns on decimal point
\usepackage{bm}% bold math
\usepackage[utf8]{inputenc}
\usepackage[T1]{fontenc}
\usepackage{float}
\usepackage{multirow}
\usepackage{stix}

\usepackage[dvipsnames]{xcolor}
 
\definecolor{C0}{HTML}{1f77b4}
\definecolor{C1}{HTML}{ff7f0e}
\definecolor{C2}{HTML}{2ca02c}
\definecolor{C3}{HTML}{d62728}
\definecolor{C4}{HTML}{9467bd}
\definecolor{C5}{HTML}{8c564b}
\definecolor{C6}{HTML}{e377c2}
\definecolor{C7}{HTML}{7f7f7f}
\definecolor{C8}{HTML}{bcbd22}
\definecolor{C9}{HTML}{17becf}

\definecolor{CB0}{rgb}{0, 0, 0}
\definecolor{CB1}{rgb}{0.4, 0.4, 0.4}
\definecolor{CB2}{rgb}{0.8, 0.8, 0.8}

\definecolor{change}{rgb}{0, 0, 0}
\definecolor{change2}{rgb}{0, 0, 0}

\begin{document}

\title[The effect of external electric fields on Si with superconducting Ga nano-precipitates]{The effect of external electric fields on silicon with superconducting gallium nano-precipitates}

% IOP author style
%\author{Brandur Thorgrimsson$^{1}$, Thomas McJunkin$^{1}$, E. R. MacQuarrie$^{1}$, S.~N. Coppersmith$^{1,2}$, M. A. Eriksson$^{1}$}

%\address{$^{1}$University of Wisconsin-Madison, Madison, WI 53706, USA}
%\address{$^{2}$School of Physics, University of New South Wales, Sydney, Australia}

%\eads{\mailto{thorgrimsson@wisc.edu}, \mailto{maeriksson@wisc.edu}}
%\eads{\mailto{thorgrimsson@wisc.edu}}

% comment out aip author style
\author{Brandur Thorgrimsson}
 \email{thorgrimsson@wisc.edu}
\author{Thomas McJunkin}
\author{E. R. MacQuarrie}
 \affiliation{University of Wisconsin-Madison, Madison, WI 53706, USA}
\author{S. N. Coppersmith}
 \affiliation{University of Wisconsin-Madison, Madison, WI 53706, USA}
 \affiliation{ University of New South Wales, Sydney, NSW 2052, Australia}%
\author{M. A. Eriksson}
 \affiliation{University of Wisconsin-Madison, Madison, WI 53706, USA}
%\email{maeriksson@wisc.edu}

\date{\today}% It is always \today, today,
             %  but any date may be explicitly specified

\begin{abstract}
%WHAT DID WE ACTUALLY DO.
Motivated by potential transformative applications of nanoelectronic circuits that incorporate superconducting elements, and by the advantages of integrating these elements in a silicon materials platform, we investigate the properties of the superconductivity of silicon ion-implanted with gallium. 
% and specifically whether superconductivity in this material system can be made voltage-gateable by using specific sample preparation methods.
Here we measure 40 different samples and explore both a variety of preparation methods (yielding both superconducting and non-superconducting samples), and the reproducibility of one of the preparation methods yielding superconducting samples.
While we find agreement with the existing literature that superconducting effects are visible in this system, we also find that this superconductivity is not influenced by voltages applied to a top gate.  The superconductivity in this materials system is not gateable for applied electric fields as large as 8 MV/cm.
%applied to these samples did not achieve a change in conductance exceeding 3\%.
We also present results of scanning transmission electron microscopy imaging of some of the same samples for which we report electronic characterization.  In agreement with the existing literature, we find that the presence of Ga precipitates is essential to the presence of a superconducting transition in these samples.  However, we also find evidence for large inhomogeneities in this system, which we discuss in connection with the lack of gateability we report here.
%The results of our electrical and transmission electron microscopy (TEM) measurements are consistent with the conduction and superconductivity being due to the presence of gallium precipitates near the $\mathrm{Si/SiO}_2$ interface as opposed to proximity-induced superconductivity in the silicon, and that these precipitates cause the material system to be macroscopically inhomogeneous which is further supported by non-zero Hall voltages measured at no external magnetic field.
%The large number of samples measured allows us to study the reproducibility, and we find that sample behavior varies greatly even among samples prepared identically and on the same die.  
% measured in the 2 samples Hall measurements are performed on.
\end{abstract}

% comment out so that article starts immediately following abstract
\maketitle

% comment out before submission, they prefer submissions in single column format even for two column journals
% from IOP latex guide:  Two-column output will begin on a new page (unlike in published double-column articles, where the two-column material starts on the same page as the abstract).
%\ioptwocol

\section{Introduction} 
The integration of superconducting and semiconducting nanostructures has the potential to enable interesting and novel quantum devices~\cite{DeFranceschi:2010p703,Larsen:2015p127001,Shim:2015p1}. %  \textbf{[Brandur, can you please add citations to the recent Veldhorst and Scappucci work?]}
One advantage of such integration is the ability to modulate superconductivity with a gate voltage, which has been demonstrated in both III-V and II-VI semiconductors, including InAs~\cite{Doh:2005p272}, InSb~\cite{Zhang:2018p74}, and HgTe\cite{Hart:2017p87}, all of which form low Schottky barrier contacts with superconductors.  Recently, the superconducting proximity effect has also been observed in compressively-strained Ge grown on relaxed SiGe~\cite{Hendrickx:2018p2835}.

%which have the Fermi level pinned near either the conduction band (InAs) or the valence band (Ge and InSb)~\cite{Gobeli:1965pA245}.
%This reduces the Schottky barrier height these materials form with metal electrodes~\cite{Dimoulas:2006p252110}, and these materials easily form ohmic contacts for the corresponding majority carrier while undoped.

The promise of these devices combined with the technological importance of silicon semiconductor devices motivates us to investigate superconductivity in silicon. Although silicon has low intrinsic spin-orbit coupling, it remains an important candidate for applications requiring spin-orbit interactions, because artificial spin-orbit coupling can be can be introduced using nanomagnet arrays, which have been proposed to enable superconducting silicon nanowires to support Majorana zero modes~\cite{Kjaergaard:2012p20503,Maurer:2018p054071,Turcotte2019preprint}.

A challenge in working with superconductivity in silicon is that the Fermi level is pinned near neither the conduction nor valence bands~\cite{Gobeli:1965pA245} making it difficult to proximitize undoped silicon.  Very high doping has been shown to circumvent this problem.  The superconducting proximity effect has been observed in silicon highly doped with boron~\cite{VanHuffelen:1992p535,VanHuffelen:1993p5170}.  And at very high boron concentrations, silicon itself becomes superconducting at temperatures below 0.35~K~\cite{Bustarret:2006p465}.

 %If the ohmic contact metal is superconducting, this low barrier allows for proximitization of the semiconductor~\cite{Hendrickx:2018p2835,vanWees:1997p469}.

An alternative approach to generating superconductivity in silicon is the implantation of Ga to form nanoprecipitates, and this method has the advantage of observed critical temperatures as high as $\sim7$~K ~\cite{Skrotzki:2010p192503,Fiedler:2011p205,Heera:2012p262602,Fischer:2013p14502,Heera:2013p83015,Heera:2015p342,Heera:2015p117217}.  This method depends on annealing of the sample after Ga implantation, which causes the Ga to diffuse, resulting in both clusters and a wide range in nominal carrier density~\cite{Fiedler:2011p205,Heera:2012p262602}. 

Encouraged by the gateability of the proximity effect in {$\mathrm{III-V}$} materials~\cite{Doh:2005p272,Larsen:2015p127001,Zhang:2018p74}, \textcolor{change}{{$\mathrm{II-VI}$} materials\cite{Hart:2017p87} and germanium\cite{Xiang:2006p208,Hendrickx:2018p2835}}, and by the high $T_\mathrm{C}$ of superconductivity in Si:Ga, we here investigate superconductivity in silicon highly doped with Ga under a wide range of annealing conditions and under large applied electric fields.
%An important open question is how repeatable and controllable is the generation of superconductivity in silicon through the annealing of silicon highly doped with gallium.
By studying two sets of samples, the first with a range of processing conditions and the second with a single set of processing conditions, we address the important open question of the controllability and repeatability of the generation of superconductivity in silicon through the annealing of silicon highly doped with gallium.  In studying the first set, we find agreement with literature results that there is a mid-range annealing temperature, in our experiments between 450 and 550~$^o$C, that produces superconductivity in Si:Ga~\cite{Skrotzki:2010p192503}.  We further test the gateability of the conductivity using a gate electrode separated from the Si:Ga by SiO$_2$, with thicknesses ranging from 30~nm to 200~nm, allowing us to apply electric fields as large as 8~MV/cm.  At temperatures corresponding to the superconducting transition edge, such fields produce very small changes in resistance, consistent with zero to within the uncertainty of the measurement.

We also find that there is a large variability in this process, such that samples on the same chip and processed identically yield different results. Such results suggest inhomogeneity, and we report Hall measurements of Si:Ga samples that reveal anisotropy consistent with such inhomogeneity.  

%Because gateability of carrier density is more likely to be successful when the carrier density is low~\cite{Gupta:2012p115006}, 
%In an attempt to find a sample preparation procedure that yields voltage-gateable superconductivity, we investigate 17 different samples with a variety of annealing conditions that yield superconducting and non-superconducting samples as well as samples that have a partial superconducting transition, where the resistance drops at ${\sim 7~\mathrm{K}}$ to a finite value. 

Based on Hall measurements of the hole carrier density in two samples with partial superconducting transitions, the carrier densities are low enough that the electric fields applied would be expected to change the carrier density by $\sim 10\%$,
greater than the observed proportional change in conductivity. Additionally, we measure non-zero Hall voltages at zero external magnetic field.
%We use Hall measurements to measure the hole carrier density in two samples with a partial superconducting transition and determine the carrier density can be changed by more than 10\% using external electric fields. 
%We compare the expected proportional change in carrier density to the observed proportional change in conductivity, and find the later to be significantly smaller
%We find the observed proportional change in conductivity to be smaller than the predicted proportional change in carrier density. 
Both observations are consistent with a nonuniform distribution of charge carriers. %and supports the hypothesis that the superconductivity occurs in gallium precipitates as opposed to proximity-induced superconductivity in the silicon.
%, a result consistent with a nonuniform distribution of charge carriers.
%SUE:  WHERE?

We investigate the reproducibility of a single preparation method by measuring the resistance of these samples at liquid helium temperature, $T_{\mathrm{LHe}}$, and we find wide variations between the resistances of 21 nominally identical samples.
For some of these devices we find that the resistance at $T_{\mathrm{LHe}}$ is more than 6 orders of magnitude lower than the room temperature resistance 
\textcolor{change}{while for others we find that their} resistance at $T_{\mathrm{LHe}}$ is only an order of magnitude lower than the room temperature resistance.
%Therefore, we find that the conductivity at $T_{\mathrm{LHe}}$ of these 21 nominally identical samples \textcolor{change}{varies by orders of magnitude}, even between devices fabricated on the same die.
%In our experiments the conductivity varies widely between samples that are nominally identical, even between devices fabricated on the same die.  
%These results are consistent with the hypothesis that the superconductivity occurs in gallium precipitates as opposed to proximity-induced superconductivity in the silicon.

We confirm the presense of gallium nano-precipitates in these samples through TEM imaging.  By comparing the temperature dependence of the resistance in samples which have had none, some, or all of their gallium nano-precipitates removed by chemical etching, we find that the supercurrent flows predominantly near the $\mathrm{Si/SiO}_2$ oxide interface.  Finally, we report that voltage-gateable superconductivity is not found in any sample. The large sample-to-sample variations of the conductivity are consistent with superconductivity that occurs in gallium precipitates as opposed to proximity-induced superconductivity in the silicon.

%We also perform electrical measurements and TEM characterization of samples in which gallium nano-precipitates have been etched away from the $\mathrm{Si/SiO}_2$ interface. By comparing the temperature dependence of the resistance in samples which have had none, some, or all of their gallium nano-precipitates removed by chemical etching, we find that the super-current flow is predominantly near the oxide interface,  
%These results are consistent with the hypothesis that the superconductivity occurs in gallium precipitates as opposed to proximity-induced superconductivity in the silicon.

The paper is organized as follows: section 2 describes in detail the various sample preparation processes and the measurement procedures.  Section 3 describes the main results including data relevant to gateability under applied electric fields and the homogeneity of the materials.  Section 4 discusses the implications of the measurement results from this large set of samples.  The appendices present additional details of the sample preparations and the experiments.

% device fabrication
\section{Methods}
\label{sec:methods}

\subsection{Sample preparation} % (fold)
\begin{figure*}[ht!]
\includegraphics[width=\textwidth]{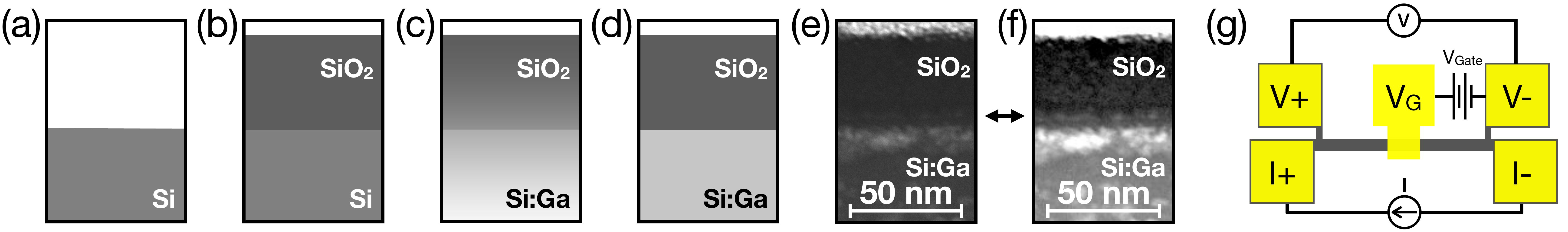}
\caption{\label{fig:methods}
(a)-(d) Schematic images of the substrate at different stages during fabrication.
(a) A (100)-oriented Si wafer with resistivity $>10~\mathrm{k\Omega cm}$ is (b) sputtered with a 30~nm thick layer of $\mathrm{SiO}_2$.
(c) Gallium is then ion implanted with energy of 80 keV and a dose of $4\times 10^{16}~\mathrm{cm}^{-2}$ at a $7^{\circ}$ angle from the sample normal. The ion beam current was kept below $0.5~\mathrm{\mu Acm}^{-2}$.
(d) Finally, the substrate undergoes a rapid thermal anneal (RTA). %process, which causes the gallium spread out and migrate towards the $\mathrm{Si/SiO}_2$ interface.
(e) Scanning transmission electron microscope (STEM) image of a sample annealed at 550~$^{\circ}$C for 30~s. The oxide thickness of this sample is $\sim40~\mathrm{nm}$. During the device fabrication (after ion implantation) this sample had 30~nm of $\mathrm{SiO}_2$ sputtered onto it in addition to the 30~nm thick $\mathrm{SiO}_2$ sputtered onto it prior to ion implantation. During the ion implantation a portion of the oxide is milled away resulting in a final oxide thickness of $\sim40~\mathrm{nm}$.
(f) The same image as in (e), but at an adjusted contrast. The arrow highlights the boundary in the oxide layer between the oxides sputtered before and after ion implantation.
(g) Schematic image of devices used to measure the electrical transport properties. Four-probe resistance measurements were performed by sourcing current from I+ to I- and measuring the voltage between V+ and V-.
An oxide separates the current channel from a top gate.
Measurements that attempt to modify the conductance of the current channel using external electric fields are performed by applying a voltage, V$_{\mathrm{Gate}}$, to the top gate, referenced to V-.}
\end{figure*}

\label{sec:sample_prep}
The sample preparation procedure is illustrated in figure~\ref{fig:methods}.  
At 300~K a nominally 30~nm layer of $\mathrm{SiO}_2$ is sputtered onto 3" float-zone n-type (100) silicon wafers (from WaferPro) with resistivity $>10~\mathrm{k\Omega cm}$ (see figure~\ref{fig:methods}(b)).  
Gallium ions with energy 80~keV and dose $4\times 10^{16}~\mathrm{cm}^{-2}$ then are implanted at a $7^{\circ}$ angle to the sample normal (figure~\ref{fig:methods}(c)); to avoid overheating the substrate, during the ion implantation the ion beam current is kept below $0.5~\mathrm{\mu Acm}^{-2}$.  
The wafers are diced into $5\times5~\mathrm{mm}^2$ dies, and individual dies undergo a rapid thermal anneal (RTA) at a variety of temperatures and annealing times (figure~\ref{fig:methods}(d)).
Table~\ref{tab1} provides a list of the 40 samples measured in this work, including descriptions of the annealing protocols used, the number of samples measured for each protocol, the types of measurements performed on each sample group\textcolor{change}{, the chemical etchant used to etch devices for each method, the ohmic contact metal, and the annealing gas used}.  The annealing conditions chosen span a range that yields both superconducting and non-superconducting samples. 
In general, superconductivity is found in samples annealed at 450-550~$^{\circ}$C for 30-900~s\textcolor{change}{, with annealing temperatures measured using a thermocouple that touches the backside of the samples during the anneal}.

\begin{table*}[t]
	\caption[]{\textbf{Summary of sample preparations and measurements.} 
	 The 40 samples reported in this paper were prepared using \textcolor{change}{5} different annealing methods, with temperatures ranging from 450~$^{\circ}$C to 650~$^{\circ}$C, and duration ranging from 15~s to 900~s. Anneals done at 450~$^{\circ}$C were performed in a forming gas atmosphere while anneals done at 550~$^{\circ}$C and above were performed in a nitrogen gas atmosphere. The table also shows the number of samples measured for each annealing method, the measurements performed on each sample group, \textcolor{change}{the chemical etchant used to etch devices for each method, the ohmic contact metal, and the annealing gas used}.
	}
	\begin{center}
	\footnotesize
	\begin{ruledtabular}
		%\begin{tabular}{|c|c|c|c|c|}
		\bgroup
		\begin{tabular}{
			|>{\centering}p{1.5cm}
			|>{\centering}p{1.6cm}
			|>{\centering}p{1.5cm}
			|>{\centering}p{1.8cm}
			|>{\centering}p{2.2cm}
			|>{\centering}p{2.2cm}
			|>{\centering}p{2.2cm}
			|>{\centering\arraybackslash}p{2.2cm}
			|}
		%\begin{tabular}{|c|c|c|c|c|c|c|c|}
			%\hline
			%\hline
			%\multirow{2}{*}{\footnotesize{Method}}& \multicolumn{2}{c|}{\footnotesize{Annealing}} &\footnotesize{Number of}& \multirow{2}{*}{\footnotesize{Measurements}}\\
			\footnotesize{Method}& \multicolumn{2}{c|}{\footnotesize{Annealing}} &\footnotesize{Number of} & \multirow{2}{*}{\footnotesize{Measurements}} & \textcolor{change}{Device Plasma} & \textcolor{change}{Ohmic} & \textcolor{change}{Annealing}\\
			\cline{2-3}
			Label &\footnotesize{Temp ($^{\circ}\mathrm{C}$)}&\footnotesize{Time (s)}&\footnotesize{Samples}& &\textcolor{change}{Etchant} & \textcolor{change}{Metal} & \textcolor{change}{Gas}\\
			\hline
			\hline
			650/30 & 650 & 30 & 5 & \footnotesize{Gating} & \textcolor{change}{Fluorine} & \textcolor{change}{Tungsten} & \textcolor{change}{N2}\\
			650/15 & 650 & 15 & 1 & \footnotesize{Gating} & \textcolor{change}{Fluorine} & \textcolor{change}{Tungsten} & \textcolor{change}{N2}\\
			550/60 & 550 & 60 & 4 & \footnotesize{Gating} & \textcolor{change}{Fluorine} & \textcolor{change}{Tungsten} & \textcolor{change}{N2}\\
			450/900 & 450 & 900 & 4 & \footnotesize{Gating} & \textcolor{change}{Fluorine} & \textcolor{change}{Tungsten} & \textcolor{change}{FG}\\
			550/30 & 550 & 30 & 3 & \begin{tabular}{@{}c@{}}\scriptsize{Location of}\\[-0.5em]\scriptsize{superconductivity}\\[-0.5em]\scriptsize{and gating}\end{tabular} & \textcolor{change}{Chlorine} & \textcolor{change}{Ti/Au} & \textcolor{change}{N2}\\
			550/30 & 550 & 30 & 21 & \footnotesize{Reproducibility} & \textcolor{change}{Chlorine} & \textcolor{change}{Ti/Au} & \textcolor{change}{N2}\\
			550/\textcolor{change}{60} & 550 & 60 & 2 & \begin{tabular}{@{}c@{}}\footnotesize{Hall}\\[-0.5em]\footnotesize{measurements}\end{tabular} & \textcolor{change}{Fluorine} & \textcolor{change}{Tungsten} & \textcolor{change}{N2}\\
			%\hline
			%\hline
		\end{tabular}
		\egroup
	\end{ruledtabular}
	\end{center}
	\label{tab1}
\end{table*}

Two oxide steps were used in the fabrication of the devices reported in this paper. 
First, all devices had 30~nm of $\mathrm{SiO}_2$ sputtered onto them before the Ga ion implantation. This oxide layer protects the silicon during the ion implantation and is partially ablated during the implantation.
Second, different amounts of additional $\mathrm{SiO}_2$ were added during subsequent processing for different devices to serve as a gate oxide.
Because we use the final oxide thickness to determine the external electric fields applied to the devices, we discuss how the final oxide thickness is determined in~\textcolor{change}{Appendix~}\ref{sec:final_oxide_thickness}.

%Lithography: 
For the devices listed in Table~\ref{tab:all} with a $100~\mathrm{nm}$ or $200~\mathrm{nm}$ additional oxide thickness, devices were etched, following the ion implantation, using a fluorine based plasma etch.
After sputtering the additional $\mathrm{SiO}_2$ gate oxide, a buffered oxide etch (BOE), hydrofluoric acid 20:1, was used to etch through the oxide over ohmic contact regions.
Tungsten was then sputtered (170~nm) over the entire wafer, lithography was performed, and the tungsten was etched using a fluorine-based plasma etch to create metallized ohmic contacts and top gates.
The wafers were diced into $5\times5~\mathrm{mm}^2$ dies, and finally individual dies underwent different RTA processes.  % in a flowing nitrogen atmosphere (1 sccm of N$_2$).
In Ref.~\onlinecite{Heera:2012p262602} \textcolor{change}{a variety of} RTA procedures were used to modify the carrier concentration; low concentrations yielded insulating samples, and high concentrations yielded superconducting samples.
\textcolor{change}{Here we make use of similar annealing procedures, which are described for each sample in Table 1.  Most of the RTA steps were performed in a flowing nitrogen atmosphere (1 sccm of N$_2$) and ranged in temperature from 550~$^{\circ}$C to 650~$^{\circ}$C and in duration from 15~s to 60~s. We also made use of a flowing forming gas (FG, 1 sccm) anneal at 450~$^{\circ}$C for 900~s anneal.}
% The RTA processes used here ranged from 550~$^{\circ}$C to 650~$^{\circ}$C in temperature and from 15~s to 60~s in duration in a flowing nitrogen atmosphere (1 sccm of N$_2$)
%While these RTA recipies are similar to what has been previously reported~\cite{Skrotzki:2010p192503,Fiedler:2011p205,Heera:2012p262602}, we note that we found that we needed to use temperatures $\sim50-100^{\circ}\mathrm{C}$ lower in order for our samples to superconduct.
% \textcolor{change}{and an anneal at 450~$^{\circ}$C for 900~s in a flowing forming gas atmosphere (1 sccm of FG).}
% An anneal at 450~$^{\circ}$C for 900~s anneal in a flowing forming gas atmosphere (1 sccm of FG) was also used. %These processes either resulted in superconducting or non-superconducting samples.

For the devices listed in Table~\ref{tab:all} with $30~\mathrm{nm}$ and $40~\mathrm{nm}$ gate oxide thickness and in figures~\ref{fig:histogram}, \ref{fig:removal}~\&~\ref{fig:variation}, after the ion implantation, the samples were diced into $5\times5~\mathrm{mm}^2$ dies and then \textcolor{change}{were annealed using} an RTA process, for 30~s at 550~$^{\circ}$C in a flowing nitrogen atmosphere (1 sccm of N$_2$).
After the annealing, multiple devices were etched for four-probe electrical measurements using a chlorine-based plasma etch. 
These devices had a $500~\mathrm{\mu m}$ long and $20~\mathrm{\mu m}$ wide channel between source and drain (schematic shown figure~\ref{fig:methods}(\textcolor{change}{g})).
BOE was used to etch through the oxide over the ohmic contact regions before the ohmics were metalized by evaporating Ti(20~nm)/Au(140~nm) using an e-beam evaporator.

\textcolor{change}{To determine the location of superconductivity within the sample heterostructures, a} few dies, reported in figure~\ref{fig:removal}, were processed further. BOE was used to remove the oxide in a $10\times20~\mathrm{\mu m}^2$ area of the current channel,
after which an additional 30~nm of $\mathrm{SiO}_2$ was sputtered and then a Ti(20~nm)/Au(280~nm) top gate was evaporated on top of the exposed region.
%This BOE either removed some or all of the gallium at the $\mathrm{Si/SiO}_2$ interface.
Finally, a BOE was used to re-expose the metalized ohmic contacts. 
\textcolor{change}{For comparison with the devices that underwent oxide removal,} Ti/Au top gates were also evaporated on some devices that had not had their oxide removed \textcolor{change}{(these devices are categorized with the "Location of superconductivity" devices in Table~\ref{tab1})}; figure~\ref{fig:methods}(\textcolor{change}{g}) shows a schematic image of these devices.
% need to mention that some or all of the gallium at the si/sio2 interface is remvoed by the hf etch.
%Cross-sectional schematic and scanning transmission electron microscope (STEM) images of these devices are in figures~\ref{fig:removal}(b)-(g) 

% subsection sample_preparation (end)

\subsection{STEM sample preparation and imaging} % (fold)
\label{sub:stem_sample_preparation}

%Scanning Tunneling Electron Microscopy (STEM) samples of devices where the gallium at the $\mathrm{Si/SiO}_2$ interface was partially removed, completely removed or not removed, using a BOE, were prepared using a FEI Helios Plasma enhanced focused ion beam scanning electron microscope. 
Scanning Transmission Electron Microscopy (STEM) samples were prepared using a Helios G4 UX DualBeam System. 
During the sample preparation an electron-beam was used to deposit a thin ($\sim200\mathrm{nm}$) platinum cover layer, a thicker (few $\mu\mathrm{m}$) cover layer was then deposited using an ion-beam.
This procedure was used because the cover layer protects the surface of the STEM sample from the ion-beam used to mill out the STEM sample.  
STEM images (not shown) of samples that had their initial cover layer deposited using an ion-beam showed significant damage $\sim$30~$\mathrm{nm}$ from the surface and into the gallium doped silicon beneath the $\mathrm{Si/SiO}_2$ interface.
%STEM samples were prepared and imaged of both devices where the $\mathrm{SiO}_2$ was and wasn't removed using BOE.
Imaging was performed using a FEI Titan 200~keV STEM using a high angle annular dark field (HAADF) image detector.  
%Cross-sectional STEM images of samples where the $\mathrm{SiO}_2$ was n gallium at the $\mathrm{Si/SiO}_2$ interface was not removed, partially removed and completely removed, using BOE were prepared.%are shown in figures~\ref{fig:removal} (e), (f) and (g) respectively.
%Cross-sectional STEM images of samples where the gallium at the $\mathrm{Si/SiO}_2$ interface was not removed, partially removed and completely removed, using BOE were prepared.%are shown in figures~\ref{fig:removal} (e), (f) and (g) respectively.

% subsection stem_sample_preparation (end)

\subsection{Electrical measurements} % (fold)
\label{sub:electrical_measurements}

Measurements were performed in an Oxford \textcolor{change}{Teslatron} cryostat, in liquid helium dewars, in a Quantum Design Physical Properties Measurement System (PPMS), and in a Janis dilution refrigerator (DR). For measurements in the Oxford Teslatron, helium dewars and Janis DR, a Keithley 2400 source meter was used to source current and measure voltage. In the PPMS a Keithly 6221 current source was used to source current and the voltage was measured using a Keithley 2182 nanovoltmeter.
For all experimental setups a Keithley 2400 source meter was used to apply top gate voltages and to measure the top gate leakage current.

%In~\cite{Heera:2012p262602} different RTA procedures were used to modify the carrier concentration, low concentrations yielded insulating samples and high concentrations yielded superconducting samples.
%Figure~\ref{fig:methods}(e) Shows a schematic image of the device layout and names each terminal of the device.  
%The resistance of samples was determined by performing a four probe measurement. % by sourcing current from I+ to I- and measuring the voltage between V+ and V-.  
%External electric fields were applied to the current channel by applying a voltage $\mathrm{V_{Gate}}$ to the top gate referenced to V-.  
As discussed above, the range of sample preparation conditions was chosen to yield both non-superconducting and superconducting samples.  
In general, non-super-conducting samples have low carrier concentrations and superconducting samples have high carrier concentrations.  
For the devices that superconduct we attempted to turn the device non-superconducting by reducing the density of carriers with an external electrical field.  
% and cause the device to stop superconducting. 
For p-type dopants, such as gallium, this means applying a positive gate voltage to reduce the number of holes.  
For non-superconducting devices we attempted to induce superconductivity by increasing the number of holes by applying a negative voltage.  
For both superconducting and non-superconducting devices, the top gate voltages were increased until breakdown voltages were reached, 
which we determined by monitoring the leakage current through the top gate while simultaneously monitoring the four probe resistance of the device.  
To compare devices with different oxide thicknesses we convert the applied top gate voltage into applied external electric field by dividing the applied top gate voltage with the oxide thickness.  We compare the conductance change at an external electric field of 1~MV/cm across multiple devices.
%The breakdown voltage was determined by monitoring the leakage current through the top gate, used to apply the voltage. 
%When the leakage current starts to exponentially increase, breakdown voltage has been reached. 
%A value for the breakdown voltage was measured for both positive and negative voltages.
%The breakdown field is then determined for superconducting devices, by dividing the positive breakdown voltage with the oxide thickness between the current channel and the top gate. 
%For normal devices the negative breakdown voltage is used.
%An example of such a leakage measurement can be found in the left inset of figure~\ref{fig:gating}. 
%The left inset of figure~\ref{fig:gating} shows the device resistance measured simultaneously to the leakage measurement shown in the right inset. 
%For each device we then determine how much the resistance changed, when the external electric field was changed from no applied field to 90\% of the breakdown field value.
%The temperature at which each sample was gated ($T_{\mathrm{Gate}}$) is shown in table~\ref{tab:all}.
%We determine the percentage change in resistance by dividing the difference in resistance at 90\% breakdown field and at no field by the average resistance measured between those two voltage values.

Hall measurements were performed in a Janis DR on samples with a square van der Pauw geometry~\cite{Keithley:2013}. The Hall voltage ($V_{\mathrm{H}}$), at an applied current ($I$), was measured as a function of external perpendicular magnetic field ($B_{\perp}$). The formula, ${n_{\mathrm{2D}}=-IB_{\perp}/(eV_{\mathrm{H}})}$, where $e$ is the elementary charge, was then used to extract the carrier type and sheet density per unit area ($n_{\mathrm{2D}}$).

% subsection electrical_measurements (end)
\section{Results}

\subsection{Superconducting transition} % (fold)
\label{sub:superconducting_transition}

\begin{figure}[t]
\includegraphics[width=8.5cm]{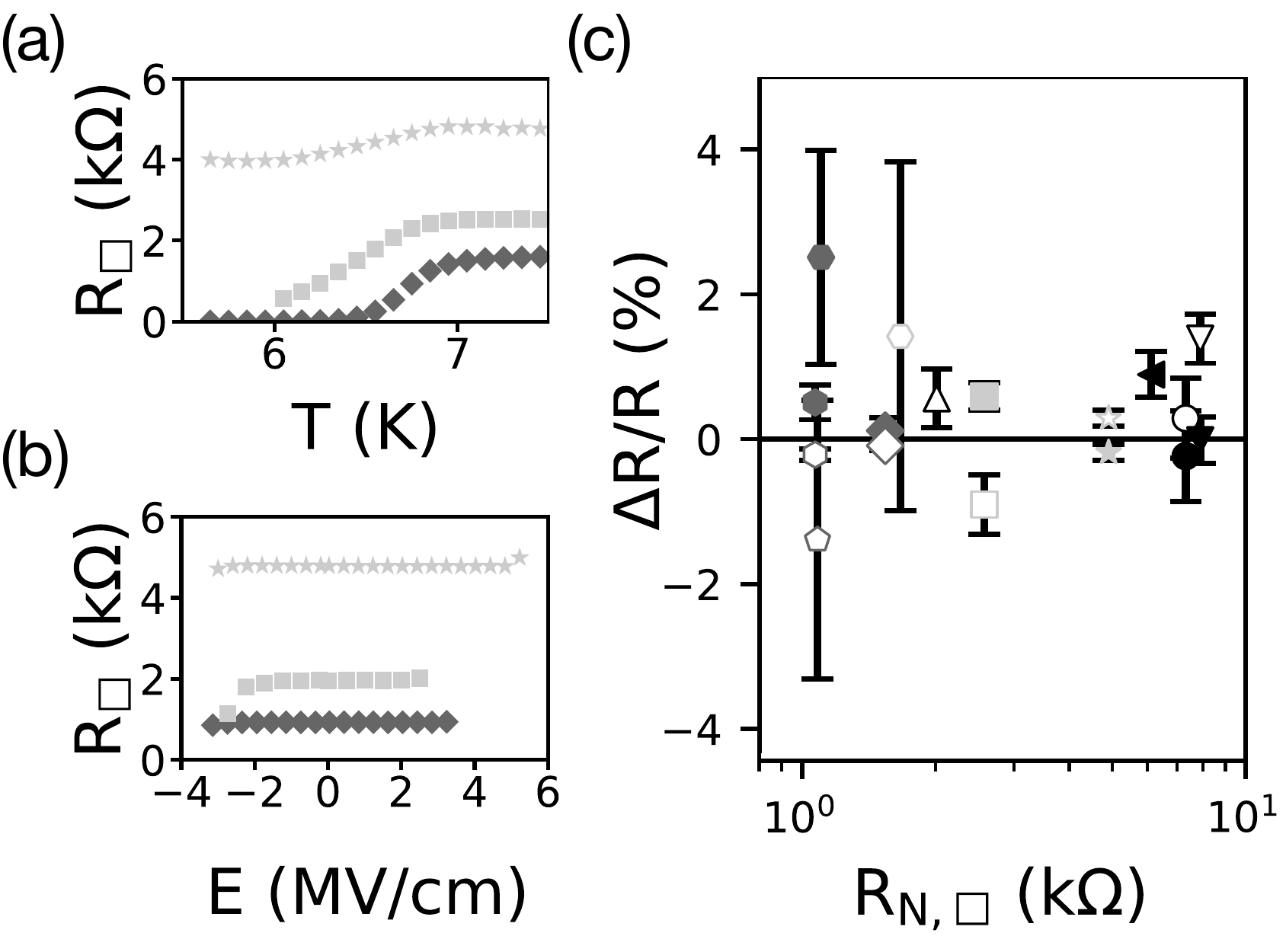}
\caption{\linespread{1.0}\selectfont{}
\label{fig:gating}
\textbf{External electric fields change the conductivity by less than 3\%.}
a) Resistance per square ($\mathrm{R_{\mdlgwhtsquare}}$) vs.\ temperature ($T$) for the three devices identified by \textcolor{CB1}{$\mdlgblkdiamond$}, \textcolor{CB2}{$\bigstar$}, \textcolor{CB2}{$\mdblksquare$} in Table~\ref{tab:all}. The devices show either a full (\textcolor{CB1}{$\mdlgblkdiamond$}) or partial (\textcolor{CB2}{$\bigstar$}, \textcolor{CB2}{$\mdblksquare$}) superconducting transition when cooled below 7~K. 
b) Resistance per square ($\mathrm{R_{\mdlgwhtsquare}}$) vs.\ external electric field ($E$) of the same devices as in (a), identified by \textcolor{CB1}{$\mdlgblkdiamond$}, \textcolor{CB2}{$\bigstar$}, \textcolor{CB2}{$\mdblksquare$} in Table~\ref{tab:all}.
These measurements were performed at the superconducting transition: 6.6~K (\textcolor{CB2}{$\mdblksquare$}) and 6.75~K (\textcolor{CB1}{$\mdlgblkdiamond$},\textcolor{CB2}{$\bigstar$}).
Less than a 3\% change is observed in the resistance of the device until after dielectric breakdown, where additional current enters the device channel through the top gate and the model of four probe resistance measurement used to determine the resistance no longer applies to the system.
c) Measured proportional change in device resistance ($\Delta\mathrm{R}/\mathrm{R}$) when an externally applied electric is changed from 0 to 1~MV/cm vs.\ the normal resistance per square ($\mathrm{R_{N,\mdlgwhtsquare}}$) of the device. 
A unique symbol represents each device. That symbol, annealing method, oxide thickness, top gate area\textcolor{change}{, the experimental conditions, the maximum applied electric field, resistance data and the temperature at which gating electric fields are applied to each sample} are summarized in Table~\ref{tab:all}. 
We use black symbols to identify non-superconducting samples, light gray symbols to identify samples with a partial superconducting transition, and dark gray symbols to identify samples with a complete superconducting transition. 
The precise geometries if the lithographically-defined Hall bars are all qualitatively the same as the device described in figure~\ref{fig:methods}(\textcolor{change}{g}) and the main text.
The value of the voltage at which the external electric field is 1~MV/cm is calculated by multiplying the electric field by the oxide thickness between the current channel and the top gate. 
For superconducting devices where we want to decrease the number of holes a positive voltage is used.
For normal devices where we want to increase the number of holes a negative voltage is used.
The plotted percentage change in resistance is calculated by dividing the difference in resistance at an external field of 1~MV/cm and no external field by the average resistance measured between those two voltage values.
The maximum change in resistance observed across the devices measured was less than 3\%.
%Table~\ref{tab:all} tabulates the experimental conditions, the maximum applied electric field, resistance data for each sample in this figure and the temperature at which gating electric fields are applied to each sample.
% regular transistors are capable of changing their source drain resistance by many orders of magnitude.
}
\end{figure}
% subsection external_electric_field_change_the_conductivity_by_less_than_5 (end)
Table~\ref{tab:all} describes the annealing protocol, oxide thickness, top gate area, \textcolor{change}{the experimental conditions, the maximum applied electric field and the raw resistance data for each sample} shown in figure~\ref{fig:gating}. The annealing protocols range from 450$^{\circ}\mathrm{C}$ to 650$^{\circ}\mathrm{C}$ in temperature and from 15~s to 900~s in duration, the oxide thickness ranges from 30~nm to 200~nm and the gate area ranges from $4~\mu \mathrm{m}^2$ to $225~\mu \mathrm{m}^2$.
The color of the data points in figure~\ref{fig:gating} identifies the different conductance behaviors of different devices when cooled below 7~K. 
Devices either show a full superconducting transition (dark gray), show a partial superconducting transition to a finite resistance value (light gray), or show no superconducting transition (black).
In general, we find that higher temperatures result in non-superconducting samples.

Figure~\ref{fig:gating}(a) shows resistance vs.\ temperature measurements for 3 different devices in the temperature range of 5.5 to 7.5~K. These three devices are identified with a \textcolor{CB1}{$\mdlgblkdiamond$}, \textcolor{CB2}{$\bigstar$} or \textcolor{CB2}{$\mdblksquare$} in Table~\ref{tab:all} and were annealed using 550$^{\circ}\mathrm{C}$ for 60~s(\textcolor{CB1}{$\mdlgblkdiamond$}, \textcolor{CB2}{$\mdblksquare$}) and 450$^{\circ}\mathrm{C}$ for 900~s(\textcolor{CB2}{$\bigstar$}).

These devices demonstrate two different behaviors observed in these experiments: the device identified by \textcolor{CB1}{$\mdlgblkdiamond$} shows a full superconducting transition, while the devices identified by \textcolor{CB2}{$\bigstar$} and \textcolor{CB2}{$\mdblksquare$} show partial superconducting transitions to (different) finite conductances. Devices with no signs of superconductivity were also measured. We note that the devices identified by \textcolor{CB1}{$\mdlgblkdiamond$} and \textcolor{CB2}{$\mdblksquare$} underwent the same annealing procedure but show different conductance behaviors.

% devices showing; a full superconducting transition are colored blue, a partial superconducting transition are colored orange, and no superconducting transition are colored black.
% subsection superconducting_transition (end)

\subsection{External electric fields change the conductivity by less than 3\%} % (fold)
\label{sub:external_electric_field_change_the_conductivity_by_less_than_5}

% Gating Results 
Figure~\ref{fig:gating}(b) shows measurements of a device resistance as a function of applied external electric field for the devices identified by~\textcolor{CB1}{$\mdlgblkdiamond$}, \textcolor{CB2}{$\bigstar$} and \textcolor{CB2}{$\mdblksquare$} in Table~\ref{tab:all}.
These measurements were performed at the superconducting transition: 6.6~K (\textcolor{CB2}{$\mdblksquare$}) and 6.75~K (\textcolor{CB1}{$\mdlgblkdiamond$},\textcolor{CB2}{$\bigstar$}).
The applied external electric field, for these three devices, are swept from; ${-3}$ to ${3~\mathrm{MV/cm}}$ (\textcolor{CB1}{$\mdlgblkdiamond$}), ${-2}$ to ${2~\mathrm{MV/cm}}$ (\textcolor{CB2}{$\mdblksquare$}), and ${-3}$ to ${4~\mathrm{MV/cm}}$ (\textcolor{CB2}{$\bigstar$}) before gate oxide breakdown is reached.
The changes in resistances are clearly negligible until the breakdown voltages.  After that voltage, current flows into the substrate from the top gate, making the four probe resistance measurement used to determine the resistances of the devices inaccurate. 
%The changes in resistances are negligible until breakdown voltage is reached, when the breakdown voltage is reached additional current flows into the devices, through the top gates, making the four probe resistance measurement used to determine the resistances of the devices inaccurate. 
The observed resistance changes in each device before breakdown are $0.85\pm 0.23~\%$ (\textcolor{CB2}{$\mdblksquare$}), $-0.34\pm 0.13~\%$ (\textcolor{CB2}{$\bigstar$}), and $0.03\pm 0.12~\%$ (\textcolor{CB1}{$\mdlgblkdiamond$}).

To summarize multiple datasets like the ones shown in figure~\ref{fig:gating}(b) we show in figure~\ref{fig:gating}(c) the proportional resistance change at 1~MV/cm vs.\ normal resistance per square, for multiple devices.
The maximum change in resistance observed across the measured devices is less than 3\%. 
%For comparison; transistors are capable of changing their resistance by many orders of magnitude.
The normal resistance per square for the devices shown ranges from 1 to 10$~\mathrm{k\Omega}$. Devices that have a full superconducting transition have lower normal resistance than devices that had a partial or no superconducting transition and devices that have a partial superconducting transition have a lower normal resistance than non-superconducting devices (with one exception, see \textcolor{CB0}{$\vartriangle$} in figure~\ref{fig:gating}(c)).
The different annealing methods used during fabrication, oxide thicknesses, and top gate areas for the different devices are described in~Table~\ref{tab:all}.
For each device, electric fields are applied until a breakdown field is reached. The maximum electric field we reach in our experiments is 8~MV/cm, and even at these higher electric fields we find no significant change in resistance across the measured devices. We note that top gates with smaller lateral dimensions are generally expected to yield higher breakdown fields, as a smaller area decreases the chance of having a defect in the oxide that may lower the breakdown field.
We also note that the data for the proportional change in resistance ($\Delta R/R_{0}$) in Table~\ref{tab:all} is the value at 90\% of the breakdown field ($E_{\mathrm{BD}}$) and not at 1~MV/cm which is shown in figure~\ref{fig:gating}.
%Table~\ref{tab:all} tabulates the experimental conditions, the maximum applied electric field and resistance data for each sample in figure~\ref{fig:gating}. 
Table~\ref{tab:all} also tabulates the temperature at which we attempted to gate each device; superconducting devices were measured at or below the superconducting transition, while \textcolor{change}{samples labelled} non-superconducting \textcolor{change}{showed no superconducting transition down to at least 4 K.}
%The devices measured here would require more than 200 times stronger electric fields than have already been applied to them to achieved a similar order of magnitude change in resistance.
%Table~\ref{tab:all} summarizes the preparation and measurements of the samples that were fabricated and characterized.  The preparation methods are listed in Table 1 of the main text.
%For samples that have a superconducting transition, the normal resistance is measured at 7.5~K. For samples without a superconducting transition the normal resistance is measured at \textcolor{change}{their} lowest temperature, typically 4-5~K.
%For devices that had no superconducting transition, gating was attempted at the lowest temperature reached.
%For devices that had a complete or partial superconducting transition, gating was attempted at or below \textcolor{change}{their} superconducting transition.

\begin{table*}[ht!]
\caption[]{\linespread{1.0}\selectfont{}
\textbf{Summary of samples fabricated and results of attempts to perform electrical gating for samples shown in figure~\ref{fig:gating}.} 
This table relates each unique symbol in figure~\ref{fig:gating} to the corresponding sample and shows the raw data for the samples we attempted to gate
%Each sample in figure~\ref{fig:gating} is given a unique symbol. 
%This table relates those symbols to the annealing method (described in table~\ref{tab1}), oxide thickness, and top gate area of each sample.
%Black symbols identify non-superconducting samples, light gray symbols identify samples with a partial superconducting transition, and dark gray symbols identify samples with a complete superconducting transition.
% Summary of samples fabricated and results of attempts to perform electrical gating.
% The table shows the raw data for the samples we attempted to gate. 
The table columns are: 
Symbol (Symb.) used to identify the sample in figure~\ref{fig:gating},
The annealing method (Meth.)\textcolor{change}{, temperature/time in degrees Celsius and seconds,} \textcolor{change}{described} in Table~\ref{tab1},
\textcolor{change}{the presence of a superconducting transition (S.C. Tr.) when cooled below 7~K, indicating if there was no transition (NO, Black symbols), a partial transition to a finite value (YES(P), light gray symbols) or a complete transition (YES, dark gray symbols).}
the minimum temperature reached during the experiment ($T_{\mathrm{min}}$), 
the resistance at \textcolor{change}{$T_{\mathrm{min}}$} ($R(T_{\mathrm{min}})$), 
the normal state resistance ($R_{\mathrm{N}}$),
the temperature at which we attempted to gate the sample ($T_{\mathrm{Gate}}$),
the gate oxide thickness ($d_{ox}$),
the external electric field breakdown value ($E_{\mathrm{BD}}$),
the area of the top gate the external electric field was applied to ($A_{\mathrm{Gate}}$),
\textcolor{change}{the average and standard deviation in resistance when applying external electric fields, ($R_{0}$) and ($\sigma_{R_{0}}$),
the change and proportional change in resistance from no external electric field to 90\% of the breakdown field value ($\Delta R$) and ($\Delta R/R_{0}$).
%and the proportional change in resistance from no external electric field to 90\% of the breakdown field value ($\Delta R/R_{0}$). 
We note that the value of $\Delta R/R_{0}$ shown in figure~\ref{fig:gating}, is the value of $\Delta R/R_{0}$ at 1~MV/cm.}
%For the presence of a superconducting transition we indicate if there was no transition (NO), a partial transition to a finite value (YES(P)) or a complete transition (YES).
For samples that have a \textcolor{change}{S.C. Tr.; $R_{\mathrm{N}}$} is measured at 7.5~K, for samples without a \textcolor{change}{S.C. Tr.; $R_{\mathrm{N}}$} is measured at \textcolor{change}{$T_{\mathrm{min}}$}.
For all samples the resistance change was less than 3\%, even when external electric fields of up to 8 MV/cm were applied.
}
\vspace{-1em}
\begin{center}
\begin{ruledtabular}
\begin{tabular}{|@{\hspace{0.5ex}}c@{\hspace{0.5ex}}|@{\hspace{0.5ex}}c@{\hspace{0.5ex}}|r|c|r|r|c|c|@{\hspace{0.5ex}}c@{\hspace{0.5ex}}|c|r|r|r|c|}
\footnotesize{Symb.} &
\footnotesize{Meth.} &
%\begin{tabular}{@{}c@{}}SC\end{tabular} & 
\footnotesize{S.C. Tr.} & 
\footnotesize{\begin{tabular}{@{}c@{}}$T_{\mathrm{min}}$\\[-0.1em](K)\end{tabular}} & 
\footnotesize{\begin{tabular}{@{}c@{}}$R(T_{\mathrm{min}})$\\[-0.1em]($k\Omega/\mdlgwhtsquare$)\end{tabular}} & 
%\begin{tabular}{@{}c@{}}$R_{\mathrm{N}}$\\[-0.1em]($\Omega$)\end{tabular} & 
\footnotesize{\begin{tabular}{@{}c@{}}$R_{\mathrm{N}}$\\[-0.1em]($k\Omega/\mdlgwhtsquare$)\end{tabular}}& 
\footnotesize{\begin{tabular}{@{}c@{}}$T_{\mathrm{Gate}}$\\[-0.1em](K)\end{tabular}} & 
\footnotesize{\begin{tabular}{@{}c@{}}$d_{ox}$\\[-0.1em](nm)\end{tabular}}& 
\footnotesize{\begin{tabular}{@{}c@{}}$E_{\mathrm{BD}}$\\[-0.1em] (MV/cm)\end{tabular}} & 
\footnotesize{\begin{tabular}{@{}c@{}}$A_{\mathrm{Gate}}$\\[-0.1em]($\mu \mathrm{m}^2$)\end{tabular}} &
%\begin{tabular}{@{}c@{}}$R_{0}$\\[-0.1em]($\Omega$)\end{tabular} & 
%\begin{tabular}{@{}c@{}}$\delta R_{0}$~\\[-0.1em]($\Omega$)~\end{tabular} & 
%\begin{tabular}{@{}c@{}}$\Delta R$~\\[-0.1em]($\Omega$)~\end{tabular} & 
\footnotesize{\begin{tabular}{@{}c@{}}$R_{0}$\\[-0.1em]($k\Omega/\mdlgwhtsquare$)\end{tabular}} & 
\footnotesize{\begin{tabular}{@{}c@{}}$\sigma_{R_{0}}$\\[-0.1em]($\Omega/\mdlgwhtsquare$)\end{tabular}} & 
\footnotesize{\begin{tabular}{@{}c@{}}$\Delta R$\\[-0.1em]($\Omega/\mdlgwhtsquare$)\end{tabular}} & 
\footnotesize{\begin{tabular}{@{}c@{}}$\Delta R/R_{0}$\\[-0.1em](\%)\end{tabular}}\\[-0.0em]
\hline
\hline
\textcolor{CB0}{$\mdlgblkcircle$} & \textcolor{change}{\footnotesize{650/30}} & \footnotesize{  NO  } & \footnotesize{4} & \footnotesize{7.33} & \footnotesize{7.33} & \footnotesize{4} & \footnotesize{100} & \footnotesize{3.39} & \footnotesize{180} & \footnotesize{7.77} & \footnotesize{50} & \footnotesize{38} & \footnotesize{0.48\%}\\[-0.2em]
\textcolor{CB0}{$\mdwhtcircle$} & \textcolor{change}{\footnotesize{650/30}} & \footnotesize{  NO  } & \footnotesize{4} & \footnotesize{7.33} & \footnotesize{7.33} & \footnotesize{4} & \footnotesize{100} & \footnotesize{1.9} & \footnotesize{225} & \footnotesize{7.76} & \footnotesize{42} & \footnotesize{37} & \footnotesize{0.48\%}\\[-0.2em]
\textcolor{CB0}{$\blacktriangledown$} & \textcolor{change}{\footnotesize{650/30}} & \footnotesize{  NO  } & \footnotesize{4} & \footnotesize{7.90} & \footnotesize{7.90} & \footnotesize{4} & \footnotesize{100} & \footnotesize{2.27} & \footnotesize{48} & \footnotesize{8.16} & \footnotesize{32} & \footnotesize{129} & \footnotesize{1.58\%}\\[-0.2em]
\textcolor{CB0}{$\triangledown$} & \textcolor{change}{\footnotesize{650/30}} & \footnotesize{  NO  } & \footnotesize{4} & \footnotesize{7.90} & \footnotesize{7.90} & \footnotesize{4} & \footnotesize{100} & \footnotesize{1.78} & \footnotesize{60} & \footnotesize{8.15} & \footnotesize{98} & \footnotesize{161} & \footnotesize{1.98\%}\\[-0.2em]
%\textcolor{CB0}{$\blacktriangle$} & \footnotesize{1} & \footnotesize{  NO  } & \footnotesize{4} & \footnotesize{2.01} & \footnotesize{2.01} & \footnotesize{3.9} & \footnotesize{100} & \footnotesize{0.38} & \footnotesize{100} & \footnotesize{2.05} & \footnotesize{4.4} & \footnotesize{7.58} & \footnotesize{0.37\%}\\[-0.1em]
\textcolor{CB0}{$\vartriangle$} & \textcolor{change}{\footnotesize{650/30}} & \footnotesize{  NO  } & \footnotesize{4} & \footnotesize{2.01} & \footnotesize{2.01} & \footnotesize{3.9} & \footnotesize{100} & \footnotesize{1.98} & \footnotesize{225} & \footnotesize{2.05} & \footnotesize{8} & \footnotesize{20} & \footnotesize{0.98\%}\\[-0.2em]
\textcolor{CB0}{$\smallblacktriangleleft$} & \textcolor{change}{\footnotesize{650/15}} & \footnotesize{  NO  } & \footnotesize{3.9} & \footnotesize{6.06} & \footnotesize{6.06} & \footnotesize{4} & \footnotesize{200} & \footnotesize{2.85} & \footnotesize{180} & \footnotesize{6.29} & \footnotesize{123} & \footnotesize{47} & \footnotesize{0.75\%}\\[-0.2em]
%\textcolor{CB0}{$\smalltriangleleft$} & \footnotesize{3} & \footnotesize{  NO  } & \footnotesize{3.9} & \footnotesize{6.06} & \footnotesize{6.06} & \footnotesize{4} & \footnotesize{200} & \footnotesize{0.64} & \footnotesize{225} & \footnotesize{6.24} & \footnotesize{18} & \footnotesize{45} & \footnotesize{0.73\%}\\[-0.1em]
%\textcolor{CB0}{$\smallblacktriangleright$} & \footnotesize{1} & \footnotesize{  NO  } & \footnotesize{4} & \footnotesize{12.10} & \footnotesize{12.10} & \footnotesize{4} & \footnotesize{200} & \footnotesize{0.4} & \footnotesize{180} & \footnotesize{12.27} & \footnotesize{49} & \footnotesize{74} & \footnotesize{0.60\%}\\[-0.1em]
%\textcolor{CB0}{$\smalltriangleright$} & \footnotesize{1} & \footnotesize{  NO  } & \footnotesize{4} & \footnotesize{12.10} & \footnotesize{12.10} & \footnotesize{4} & \footnotesize{200} & \footnotesize{0.73} & \footnotesize{225} & \footnotesize{12.25} & \footnotesize{103} & \footnotesize{117} & \footnotesize{0.96\%}\\[-0.1em]
\textcolor{CB2}{$\mdblksquare$} & \textcolor{change}{\footnotesize{450/900}} & \footnotesize{  YES(P)  } & \footnotesize{6} & \footnotesize{0.49} & \footnotesize{2.52} & \footnotesize{6.6} & \footnotesize{200} & \footnotesize{2.3} & \footnotesize{4} & \footnotesize{1.96} & \footnotesize{4.35} & \footnotesize{17} & \footnotesize{0.85\%}\\[-0.2em]
\textcolor{CB2}{$\mdlgwhtsquare$} & \textcolor{change}{\footnotesize{450/900}} & \footnotesize{  YES(P)  } & \footnotesize{6} & \footnotesize{0.49} & \footnotesize{2.52} & \footnotesize{6.6} & \footnotesize{200} & \footnotesize{6} & \footnotesize{16} & \footnotesize{1.96} & \footnotesize{5.6} & \footnotesize{-1.39} & \footnotesize{-0.07\%}\\[-0.2em]
\textcolor{CB1}{$\mdlgblkdiamond$} & \textcolor{change}{\footnotesize{450/900}} & \footnotesize{  YES  } & \footnotesize{5.6} & \footnotesize{0.00} & \footnotesize{1.50} & \footnotesize{6.75} & \footnotesize{200} & \footnotesize{2.52} & \footnotesize{4} & \footnotesize{0.92} & \footnotesize{1.23} & \footnotesize{0.32} & \footnotesize{0.03\%}\\[-0.2em]
\textcolor{CB1}{$\mdlgwhtdiamond$} & \textcolor{change}{\footnotesize{450/900}} & \footnotesize{  YES  } & \footnotesize{5.6} & \footnotesize{0.00} & \footnotesize{1.50} & \footnotesize{6.75} & \footnotesize{200} & \footnotesize{5.19} & \footnotesize{16} & \footnotesize{1.00} & \footnotesize{0.6} & \footnotesize{-0.25} & \footnotesize{-0.03\%}\\[-0.2em]
\textcolor{CB2}{$\bigstar$} & \textcolor{change}{\footnotesize{550/60}} & \footnotesize{  YES(P)  } & \footnotesize{5.6} & \footnotesize{4.00} & \footnotesize{4.82} & \footnotesize{6.75} & \footnotesize{200} & \footnotesize{5.21} & \footnotesize{8} & \footnotesize{4.79} & \footnotesize{5.9} & \footnotesize{-16} & \footnotesize{-0.34\%}\\[-0.2em]
\textcolor{CB2}{$\bigwhitestar$} & \textcolor{change}{\footnotesize{550/60}} & \footnotesize{  YES(P)  } & \footnotesize{5.6} & \footnotesize{4.00} & \footnotesize{4.82} & \footnotesize{6.75} & \footnotesize{200} & \footnotesize{3.62} & \footnotesize{16} & \footnotesize{4.80} & \footnotesize{6.26} & \footnotesize{12} & \footnotesize{0.24\%}\\[-0.2em]
\textcolor{CB1}{$\varhexagonblack$} & \textcolor{change}{\footnotesize{550/60}} & \footnotesize{  YES  } & \footnotesize{5.6} & \footnotesize{0.03} & \footnotesize{1.05} & \footnotesize{6.75} & \footnotesize{200} & \footnotesize{2.81} & \footnotesize{36} & \footnotesize{0.53} & \footnotesize{1.2} & \footnotesize{2.69} & \footnotesize{0.51\%}\\[-0.2em]
\textcolor{CB1}{$\varhexagon$} & \textcolor{change}{\footnotesize{550/60}} & \footnotesize{  YES  } & \footnotesize{5.6} & \footnotesize{0.03} & \footnotesize{1.05} & \footnotesize{6.75} & \footnotesize{200} & \footnotesize{2.59} & \footnotesize{64} & \footnotesize{0.52} & \footnotesize{0.43} & \footnotesize{0.19} & \footnotesize{0.04\%}\\[-0.2em]
%\textcolor{CB2}{$\pentagonblack$} & \footnotesize{4} & \footnotesize{  YES(P)  } & \footnotesize{5.1} & \footnotesize{0.21} & \footnotesize{1.03} & \footnotesize{5} & \footnotesize{200} & \footnotesize{0.8} & \footnotesize{22500} & \footnotesize{0.10} & \footnotesize{1.81} & \footnotesize{-1.82} & \footnotesize{-1.78\%}\\[-0.1em]
\textcolor{CB1}{$\pentagon$} & \textcolor{change}{\footnotesize{550/30}} & \footnotesize{  YES  } & \footnotesize{2.5} & \footnotesize{5.0E-06} & \footnotesize{1.04} & \footnotesize{4} & \footnotesize{40} & \footnotesize{3.6} & \footnotesize{100} & \footnotesize{2.5E-4} & \footnotesize{6.5E-2} & \footnotesize{2.3E-3} & \footnotesize{0.87\%}\\[-0.2em]
\textcolor{CB1}{$\hexagonblack$} & \textcolor{change}{\footnotesize{550/30}} & \footnotesize{  YES } & \footnotesize{2.5} & \footnotesize{2.0E-06} & \footnotesize{1.06} & \footnotesize{5} & \footnotesize{40} & \footnotesize{5} & \footnotesize{100} & \footnotesize{2.6E-4} & \footnotesize{4.5E-3} & \footnotesize{4.5E-3} & \footnotesize{1.69\%}\\[-0.2em]
\textcolor{CB2}{$\hexagon$} & \textcolor{change}{\footnotesize{550/30}} & \footnotesize{ YES(P) } & \footnotesize{4.5} & \footnotesize{0.15} & \footnotesize{1.60} & \footnotesize{5} & \footnotesize{30} & \footnotesize{8} & \footnotesize{100} & \footnotesize{0.95} & \footnotesize{16} & \footnotesize{19.75} & \footnotesize{2.09\%}\\[-0.2em]
\end{tabular}
\end{ruledtabular}
\end{center}
\label{tab:all}
\end{table*}

\subsection{Resistance of nominally identical samples varies by \textcolor{change}{many} orders of magnitude below $T_{C}$} % (fold)
\begin{figure}[t]
\includegraphics[width=8.5cm]{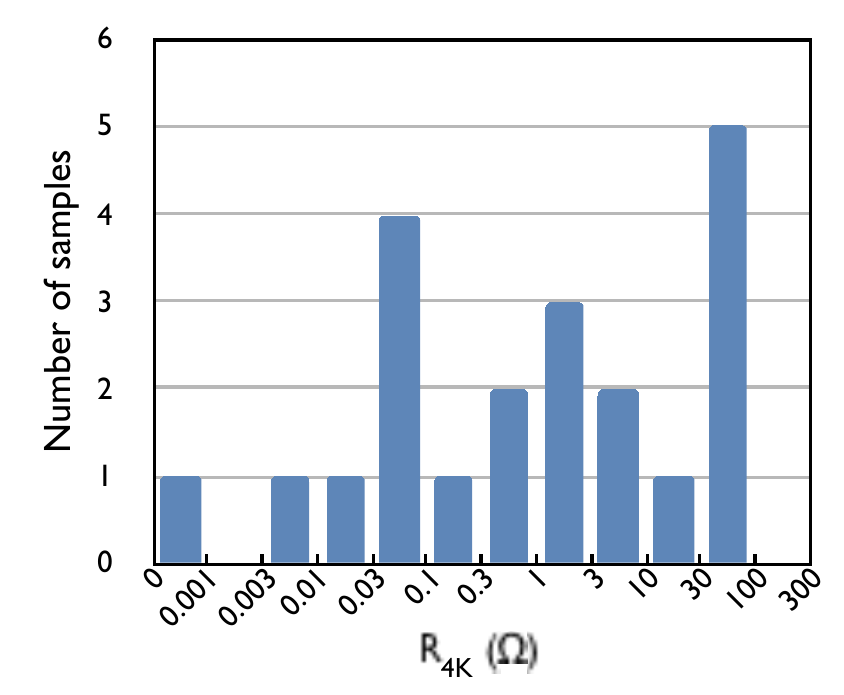}
\caption{\label{fig:histogram}
\textbf{The electrical transport properties of nominally identical devices can vary by \textcolor{change}{many} orders of magnitude.} 
A histogram of the resistance at $T=4.2$~K, $R_{4K}$, of 21 nominally identical devices, annealed at 550$^{\circ}\mathrm{C}$ for 30~s.  The labels are the resistance boundaries between each column. The room temperature resistance\textcolor{change}{s} of these devices are ${984\pm 108~\Omega}$.
Every device showed a superconducting transition when cooled below 7~K (see, e.g., figure~\ref{fig:gating}(a)), with conductances below the transition at least an order of magnitude greater than the normal state conductance.
%Even though these devices are nominally identical, their $R_{4K}$  varied by more than 5 orders of magnitude, even in devices fabricated on the same ${5\times5~\mathrm{mm}^2}$ die. 
%The four probe electrical sheet resistance was measured for 21 nominally identical devices in liquid helium.  however the residual resitance below the superconducting transition of these samples varied by more than 5 order of magnitude. 
}
\end{figure}

\label{sub:resistance_varies_by_5_orders_of_magnitude_below_t__c}
Through our gating attempts we found that nominally identical devices had a greater than expected ($>$few 10\%) variation in resistance below $T_{C}$. To determine the extent of this variation we
prepare 21 nominally identical devices annealed at 550$^{\circ}\mathrm{C}$ for 30~s and measure them in liquid helium at 4.2~K.
A resistance drop is observed in all devices when they are cooled below the superconducting transition temperature, $T_{C}\approx 6.7~\mathrm{K}$ (like the example shown in figure~\ref{fig:gating}(a)).
Figure~\ref{fig:histogram} shows a histogram of the four probe resistance $R_{4K}$ at a temperature $T=4.2$~K of these 21 nominally identical devices. 
%We call the resistance measured below $T_{C}$ a residual resistance. 
As shown in figure~\ref{fig:histogram}, the resistance at $T=4.2$~K \textcolor{change}{varies by orders of magnitude}, from <1~$\mathrm{m\Omega}$ to $\sim$90$~\Omega$. \textcolor{change}{We note that such variations are not unexpected in the presence of inhomogeneity in samples consisting of a percolative superconducting network, as has been suggested (Ref. \onlinecite{Heera:2016p105203}) could be present in this materials system.} We \textcolor{change}{also} note that a quarter of the devices have a resistance below ${0.06~\Omega}$, another quarter a resistance between ${0.06}$ and  ${1~\Omega}$, a third quarter a resistance between ${1}$ and  ${30~\Omega}$, and a quarter of the devices have a resistance greater than ${30~\Omega}$.
\textcolor{change}{This variation is observed across nominally identical devices, some located on the same ${5\times5~\mathrm{mm}^2}$ die. }
%This variation is observed even though these devices are nominally identical, some even located on the same ${5\times5~\mathrm{mm}^2}$ die. 
The room temperature resistances of these 21 nominally identical devices are ${984\pm 108~\Omega}$ \textcolor{change}{and a similar ($\sim\pm 10\%$) variation was observed in the resistances of these devices just before they were cooled below the superconducting transition}.
We note that even for the highest resistances at $T=4.2$~K, the resistance drop observed when the devices are cooled below $T_{C}$ is greater than an order of magnitude.

% subsection resistance_varies_by_5_orders_of_magnitude_below_t__c (end)

\subsection{Structural and transport studies of samples with different surface treatments.} % (fold)
%\subsection{Superconductivity is due to Ga at the $\mathrm{Si/SiO}_2$ interface} % (fold)
\label{sub:stem_images}

\begin{figure*}[t]
\includegraphics[width=\textwidth]{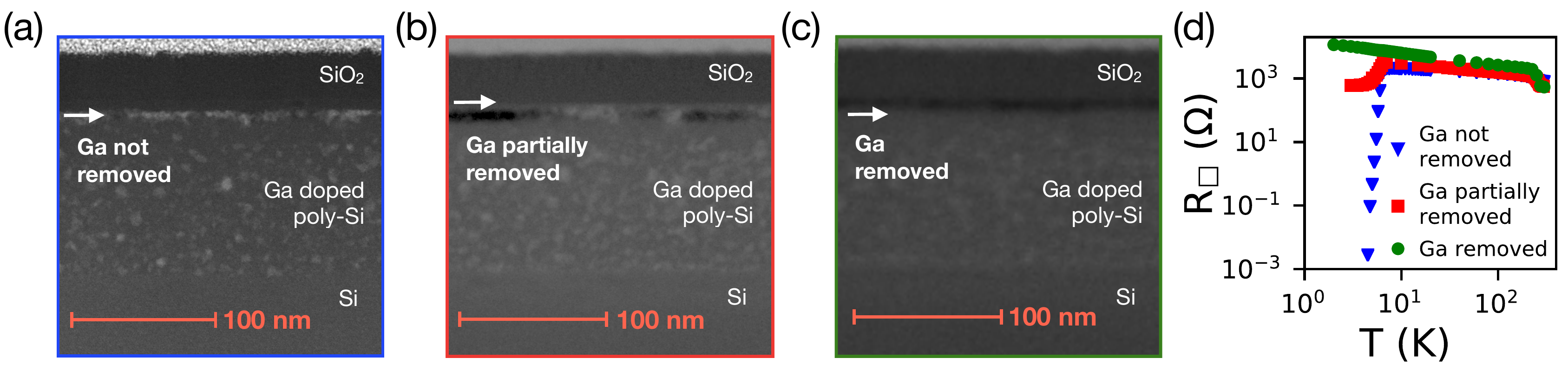}
\caption{\label{fig:removal} 
The electrical transport properties are dominated by gallium precipitates located at the $\mathrm{Si/SiO}_2$ interface. 
(a) STEM image of a device not modified with respect to what is described in figure~\ref{fig:methods}, i.e., no BOE etching perormed. 
Energy dispersive x-ray spectroscopy (EDX) measurements (not shown) confirm that both the bright spots at the $\mathrm{Si/SiO}_2$ interface and the speckles below the interface are gallium. 
%Note that the area where the gallium concentration is modified is only 5\% of the total channel area.   
(b) STEM image of a device on a die where a BOE etch resulted in a partial removal of the gallium precipitates at the $\mathrm{Si/SiO}_2$ interface.
(c) STEM image of a device on the same die as the image shown in (b), but where the same BOE etch resulted in a complete removal of the gallium precipitates at the $\mathrm{Si/SiO}_2$ interface.
%These devices were annealed at 550$^{\circ}\mathrm{C}$ for 30~s.  
%After the gallium removal 30~nm of $\mathrm{SiO}_2$ were sputtered on top of the device before a Ti/Au top gate was added. 
(d) Resistance vs.\ temperature for all 3 devices. 
%Depending on the amount of gallium precipitates present at the $\mathrm{Si/SiO}_2$ interface different behaviors are observed when the devices are cooled below 7~K.
%(Ga peak not removed, blue), a resistance drop to a fixed value (Ga peak partially removed, red) or a continued rise in resistance (Ga peak removed, green).  
\textcolor{change2}{These three devices, with different amounts of gallium precipitates present at the $\mathrm{Si/SiO}_2$ interface, show different behaviors when cooled below 7 K.}
For an unetched device (blue triangles), a full superconducting transition is observed; 
for a device where the gallium precipitates at the $\mathrm{Si/SiO}_2$ interface were partially removed using BOE (red squares, same device as a STEM image is shown of in (b)), a resistance drop to a finite value is observed;
and for a device where the gallium precipitates at the $\mathrm{Si/SiO}_2$ interface were completely removed using the same BOE etch (green circles, same device as a STEM image is shown of in (c)), a continued rise in resistance is observed.
}
\end{figure*}

Previous work has demonstrated that during the RTA process the implanted gallium forms precipitates~\cite{Nygren:1988p405} and migrates towards the $\mathrm{Si/SiO}_2$ interface~\cite{Skrotzki:2010p192503}. 
\textcolor{change}{To connect the results reported above to these same structural features, we study 3 devices, fabricated on two dies, which have different densities of Ga precipitates at the $\mathrm{Si/SiO}_2$ interface.}
%To determine how these precipitates affect the tranport properties of this material system, we study 3 devices fabricated on two dies. 
The \textcolor{change}{first} STEM image, shown in figure~\ref{fig:removal}(a), is of a die that did not undergo a BOE etch and shows an increased concentration of gallium precipitates at the $\mathrm{Si/SiO}_2$ interface compared to the surrounding material.
Figure~\ref{fig:removal}(b) \& (c) show STEM images of \textcolor{change}{the other} two devices. Both are from the same die and underwent the same BOE etch. 
For the \textcolor{change}{second} device, shown in figure~\ref{fig:removal}(b), this etch resulted in partial removal of the gallium precipitates at the $\mathrm{Si/SiO}_2$ interface;
for the \textcolor{change}{third} device, shown in figure~\ref{fig:removal}(c), this etch resulted in a complete removal of the gallium precipitates at the $\mathrm{Si/SiO}_2$ interface.
Electrical measurements were performed on these devices before STEM samples were prepared and imaged.
Gallium precipitates are apparent in these STEM images, which also show mono-crystalline undoped silicon (at the bottom), polycrystalline silicon with gallium nano-precipitates (in the middle), and increased or removed gallium nano-precipitates at a $\mathrm{Si/SiO}_2$ interface, and $\mathrm{SiO}_2$ (near the top). \textcolor{change}{At the very top of the images a glimpse of either a platinum cover layer, figure~\ref{fig:removal}(a), or a titanium sticking layer for the top gate, figures~\ref{fig:removal}(b) and (c), can be seen}.

Figure~\ref{fig:removal}(d)  shows resistance vs.\ temperature measurements for these three different devices in the temperature range $2$ to $300~\mathrm{K}$. The samples were each annealed at 550$^{\circ}\mathrm{C}$ for 30~s, and as can be seen in figure~\ref{fig:removal}(d), such samples show different behaviors as they are cooled below 7~K.
A device where gallium precipitates at the $\mathrm{Si/SiO}_2$ interface were partially removed (red squares) shows a resistance drop to a fixed value;
for the device where all the gallium precipitates at the $\mathrm{Si/SiO}_2$ interface were removed (green circles) the resistance continues to rise; 
and the unetched device (blue triangles) shows a full superconducting transition.
These measurements show that the superconductivity observed in these samples is \textcolor{change2}{observed only in the presence of} the gallium precipitate\textcolor{change2}{s} at the $\mathrm{Si/SiO}_2$ interface, 
%as partially removing the gallium precipitates from this layer results in a partial transition below $\mathrm{T_C}$ to a finite resistance value, 
\textcolor{change2}{as completely removing them results in a continued resistance rise upon decreasing temperature, consistent with previous observations in this material system~\cite{Skrotzki:2010p192503}.}

\subsection{Hall measurement data} % (fold)
\label{sec:hall_measurement_data}

\begin{figure}[t]
\includegraphics[width=8.5cm]{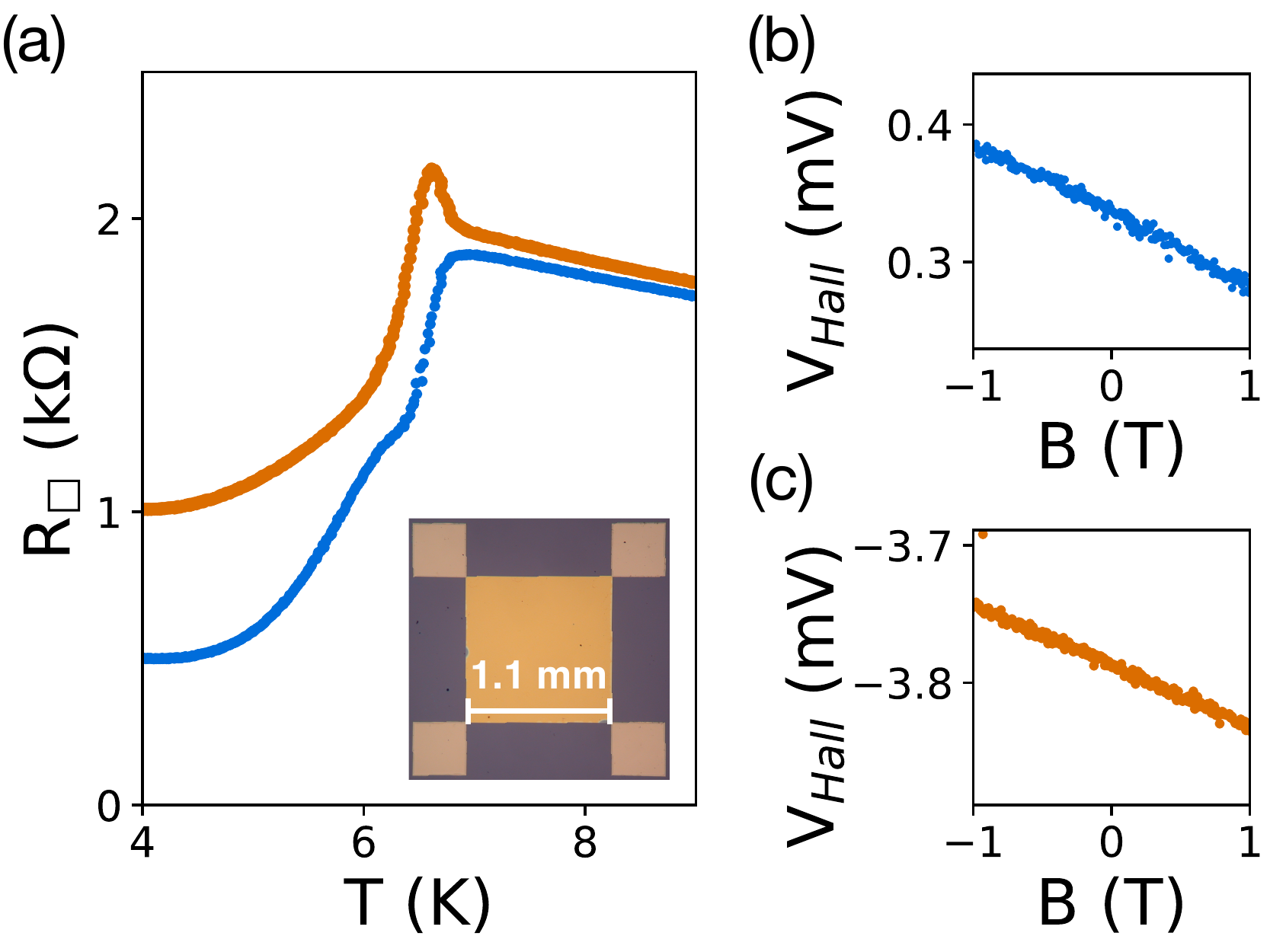}
\caption{\label{fig:hall}
\textcolor{change2}{a)} Resistance per square ($\mathrm{R_{\mdlgwhtsquare}}$) vs.\ temperature ($T$) for the two devices on which Hall measurements were performed: device 1 (\textcolor{change}{blue}) and device 2 (\textcolor{change}{orange}).
\textcolor{change2}{\emph{Inset:} Optical image of a device lithographically identical to the ones on which the Hall mesurements were performed.}
\textcolor{change2}{b)-c)} Hall (transverse) voltage (V$_{\mathrm{Hall}}$) vs.\ magnetic field (B) for device 1(2) at 8~K.  A \textcolor{change}{1}0~$\mu$A longitudinal current was sourced and a longitudinal voltage of  12.52(12.89)~mV was measured at 0~T.
The negative Hall slope shows that holes are the majority carrier. The carrier density of device\textcolor{change2}{s 1 and 2 were} calculated \textcolor{change2}{to be} (assuming the samples are uniform) $1.2\times 10^{14}$ \textcolor{change2}{and} $1.4\times 10^{14}~\mathrm{cm}^{-2}$\textcolor{change2}{, respectively}.
%The non-zero Hall voltages measured at zero external magnetic field \textcolor{change}{suggest} that the devices created in this material system are inhomogeneous.
}
\end{figure}

Figure~\ref{fig:hall}\textcolor{change2}{(a)} shows resistance vs.\ temperature measurements and Hall measurements for two devices. Both samples show a partial superconducting transition to a finite resistance value when cooled below 6.5~K. 
%\textcolor{change}{The insets of} 
Figures~\ref{fig:hall}\textcolor{change2}{(b) \& (c)} show the Hall measurement data for two devices on which Hall measurements were performed at 8~K. Both samples were annealed at 550$^{\circ}\mathrm{C}$ for 60~s.
For the first (second) sample a $-102~\mathrm{\mu V}$ ($-90~\mathrm{\mu V}$) change in Hall voltage is observed when the perpendicular magnetic field is changed from ${-1}$ to ${1~\mathrm{T}}$ when sourcing a $10~\mathrm{\mu A}$ longitudinal current. 
A longitudinal resistance per square of $1.80~\mathrm{k\Omega}$ ($1.86~\mathrm{k\Omega}$) was measured at 8~K.
The negative Hall slope shows that holes are the majority carrier. If the sample is a uniform sheet then the carrier density of device 1(2) can be calculated to be $1.2\times 10^{14}$($1.4\times 10^{14}$)$~\mathrm{cm}^{-2}$. 
However, we note that the Hall voltage at zero external magnetic field is offset from 0~V by 0.35~mV (see \textcolor{change2}{figure~\ref{fig:hall}(b)}) and -3.8~mV (see \textcolor{change2}{figure~\ref{fig:hall}(c)}), respectively, and that if a sample being measured with a Hall measurement is a uniform sheet then the Hall voltage should be 0~V at zero external magnetic field. 
The lack of the expected symmetry of these Hall measurements about zero magnetic field \textcolor{change}{could} therefore \textcolor{change}{suggest} that the devices created in this material system are not uniform.
% \textcolor{change}{or it could be a sign of ohmic contact imperfections}.

\section{Discussion}

\subsection{Conductance changes due to external electric fields} % (fold)
\label{sub:conductance_changes_due_to_external_electric_fields}

The maximum change in resistance observed across the devices measured is less than 3\% with gate-applied electric fields reaching 8~MV/cm.
It is interesting to consider how much a 8~MV/cm electric field changes the hole carrier density in our devices, especially when compared to the hole carrier sheet density $h=1.2\times 10^{14}~\mathrm{cm}^{-2}$ and $h=1.4\times 10^{14}~\mathrm{cm}^{-2}$ measured by Hall measurements.
If we assume a parallel capacitor model between the gate and channel in our devices, we calculate an expected change in hole carrier sheet density ($\delta h$) due to an external electric field ($E$) of $\delta h=\varepsilon_{r}\varepsilon_{0}E/e$, where $\varepsilon_{0}$ is the permittivity in vacuum, $\varepsilon_{r}=3.9$ is the relative permittivity of $\mathrm{SiO}_{2}$ and $e$ is the electron charge.
For an electric field of 8~MV/cm we get a change in hole carrier sheet density $\delta h=1.7\times 10^{13}~\mathrm{cm}^{-2}$ which is ${\sim13\%}$ of the hole carrier sheet density measured by the Hall measurements.
%The devices measured here would require more than 200 times stronger electric fields than have already been applied to them to achieved a similar order of magnitude change in resistance.

We can also emphasize how small of a change 3\% conductance change is by noting that a MOSFET with a $d_{ox}~=~30~\mathrm{nm}$ thick oxide and a $d_{d}~=~1~\mathrm{nm}$ thick depletion layer has a sub-threshold slope of $62~\mathrm{mV/dec}$ at 10~K (note that an idealized MOSFET would have $d_{ox}<<d_{d}$ and the sub-threshold slope approaches $2~\mathrm{mV/dec}$ at 10~K~\cite{Beckers:2018p3617}).
A 62~mV change in top gate voltage of such a transistor would change the source drain resistance by an order of magnitude. %(The applied external electric field to the current channel in this case would be 20~KV/cm).
If we assume we could change the conductance of our devices by 3\% with an 8~MV/cm field, then to change the resistance of the Si:Ga devices reported here by an order of magnitude electric fields more than 100 times stronger than those we have already applied would be required.
We note that the highest reported breakdown fields we could find for $\mathrm{SiO}_2$ films is around 27~MV/cm~\cite{Usui:2013p7660}, less than 4 times higher than the 8~MV/cm breakdown field measured here. 
Since strong external electric fields (of up to 8~MV/cm) are only able to change the conductivity of these samples by less than 3\% and carrier density by ${\sim13\%}$, this material system is not promising for voltage-gateable superconductivity.

% subsection conductance_changes_due_to_external_electric_fields (end)

\subsection{Resistance variations below $T_{C}$ and inhomogeneity} % (fold)
\label{sub:resistance_variations_below_t__c}

% subsection resistance_variations_below_t__c (end)
We note that the STEM images in figure~\ref{fig:removal}(c) show that the devices reported here are inhomogeneous on the scale of \textcolor{change}{tens} of nm (the size of the gallium nano-precipitates). \textcolor{change}{Substantial} inhomogeneity is also observed on a macroscopic scale, as the resistance of nominally identical samples varies by \textcolor{change}{many} orders of magnitude below the superconducting transition. %, and \textcolor{change}{suggested by our measurements of} non-zero Hall voltages at zero external magnetic field.
One possible explanation for this behavior is that the gallium nano-precipitates at the $\mathrm{Si/SiO}_2$ interface are superconducting but that \textcolor{change}{the conductance between these precipitates, or a precipitate network, varies. This could either be caused by varying doping levels of the material between the precipitates or by a variation in the density of the precipitates.}

%It is plausible that such inhomogeneity could exhibit large sample-to-sample variations.
%Such as a varying residual resistance below the superconducting transition of the precipitates or non-zero Hall voltages at zero external magnetic field. 

%In this explanation, the highest values measured for $\mathrm{I^*}$ would match the critical current $\mathrm{I_C}$.
%In order to test this explanation we used BOE to remove some of the gallium at the $\mathrm{Si/SiO}_2$ interface. This should significantly reduce the conductivity of the volume between the remaining gallium, but we should still see a resistance drop when the device is cooled below $\mathrm{T_C}$, albeit to a finite value.
%The data shown in figure~\ref{fig:removal}(d), where the gallium was partially removed from the $\mathrm{Si/SiO}_2$ interface and the resistance drops to a finite value when the sample is cooled below $\mathrm{T_C}$ supports this explanation.

\section{Conclusion}
Motivated by the gateability of the proximity effect in III-V materials~\cite{Doh:2005p272,Larsen:2015p127001,Zhang:2018p74}, \textcolor{change}{II-VI materials\cite{Hart:2017p87} and germanium\cite{Xiang:2006p208,Hendrickx:2018p2835}}, we sought to determine whether the well-established superconductivity in silicon doped with gallium~\cite{Skrotzki:2010p192503,Fiedler:2011p205,Heera:2012p262602,Fischer:2013p14502,Heera:2013p83015,Heera:2015p342,Heera:2015p117217} is also gateable.
With a goal of identifying a method for establishing gateable superconductivity in gallium-doped silicon, we prepared and measured  a series of samples with different annealing protocols and measured their conductivity.
% Conclusions
We saw less than a 3\% change in device conductance when applying external electric fields, even when applying fields of up to 8 MV/cm. %For comparison, the conductance of transistors, measured at cryogenic temperatures, can be changed by orders of magnitude by using orders of magnitude smaller electric fields.

In nominally identical devices measured at liquid helium temperatures (below the superconducting transition), we measured a variability of \textcolor{change}{many} orders of magnitude in the resistance.
Such variation is even found in nominally identical devices located on the same die, indicating that this material system is inhomogenous on a macroscopic scale. 

We showed that the superconductivity is only present in the gallium rich layer at the $\mathrm{Si/SiO}_2$ interface in these samples: 
%partially removing the gallium in this layer still resulted in a resistance drop at $\mathrm{T_C}$, but this drop was to a non-zero resistance at $T=4.2$~K. 
Completely removing the gallium in this layer resulted in a resistance that continued to rise as the temperature was lowered below $\mathrm{T_C}$. 
%The conductivity of the device where gallium was partially removed from the $\mathrm{Si/SiO}_2$ interface also changed less than 3\% when external electrical fields were applied.
%This demonstrates that we are neither able to use external electric fields to change the conductivity of the gallium rich layer at the $\mathrm{Si/SiO}_2$ interface nor of the underlying gallium doped poly-silicon layer, in a way such that it more than minutely effects the measured conductance.

The less than 3\% change in device conductance when applying external electric fields and the \textcolor{change}{many} order-of-magnitude variation in device resistances at $T=4.2$~K below the superconducting transition leads us to conclude that this material system is not a promising system for voltage-gateable superconductivity.

\section*{Acknowledgements}

The authors acknowledge helpful discussions with J.\ C.\ McCallum, R.\ F.\ McDermott, D.\ E.\ Savage, Viktoriia Kornich, Mark Friesen, M.\ G.\ Vavilov and A.\ Levchenko. We acknowledge Song Jin for allowing us use of his PPMS. \textcolor{change2}{The data that support the findings of this study are available from the corresponding author upon reasonable request.}

This work was supported in part by NSF EAGER under Grant No.\ DMR-1743986 and the Vannevar Bush Faculty Fellowship program sponsored by the Basic Research Office of the Assistant Secretary of Defense for Research and Engineering and funded by the Office of Naval Research through Grant No. N00014-15-1-0029, and by the Army Research Office (W911NF-17-1-0274). The views and conclusions contained in this document are those of the authors and should not be interpreted as representing the official policies, either expressed or implied, of the Army Research Office (ARO), or the U.S. Government. The U.S. Government is authorized to reproduce and distribute reprints for Government purposes notwithstanding any copyright notation herein.  The authors gratefully acknowledge use of facilities and instrumentation at the UW-Madison Wisconsin Centers for Nanoscale Technology (wcnt.wisc.edu) partially supported by the NSF through the University of Wisconsin Materials Research Science and Engineering Center (DMR-1720415).

\section*{Appendices}
\setcounter{section}{1}
\appendix

\section{Correlation between resistance at $T=4.2$~K below $T_{C}$ and $I_{C}$}

\begin{figure}[ht!]
\includegraphics[width=8.5cm]{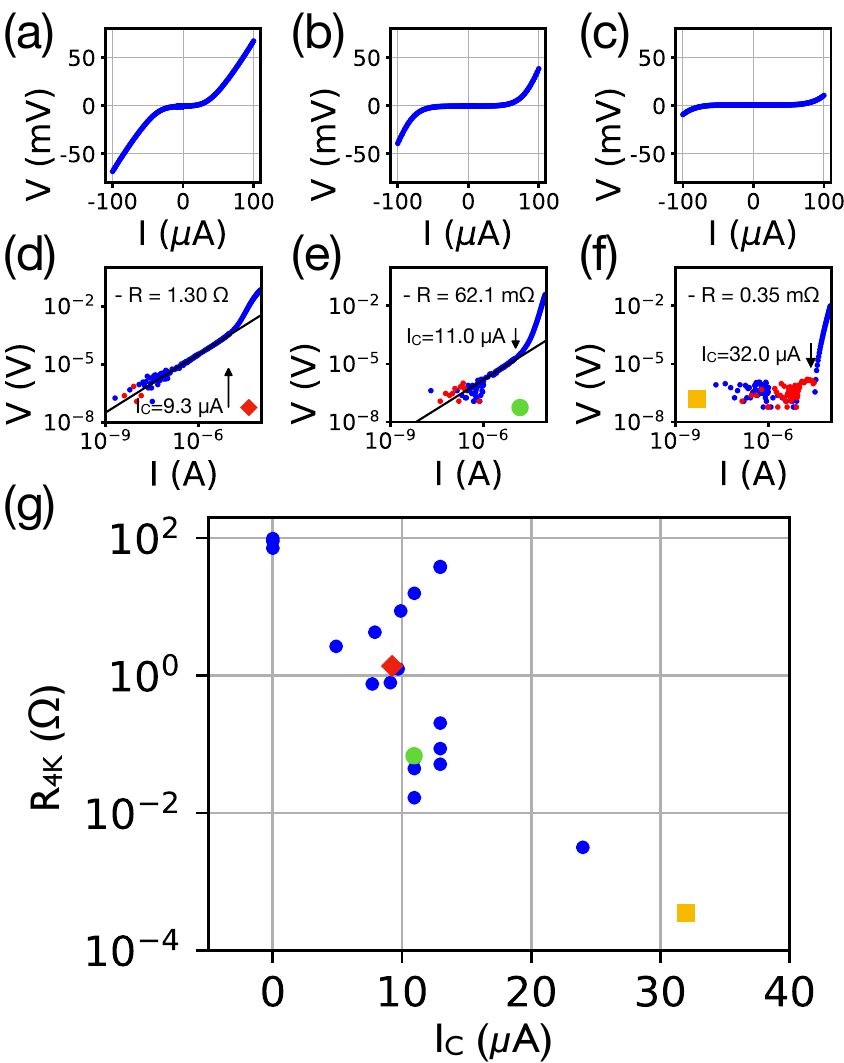}
\caption{\label{fig:variation}
\textcolor{change}{Correlation between resistance at $T=4.2$~K and critical current for the nominally identical devices, annealed at $550^{\circ}\mathrm{C}$ for 30~s, corresponding to figure~\ref{fig:histogram} above.}
% Correlation between resistance and critical current for different nominally identical devices, for the devices corresponding to results from Fig. 3 in the main text.
%The electrical transport properties of nominally identical devices can vary by \textcolor{change}{many} orders of magnitude, while their critical current varies by an order of magnitude.
%The devices reported in this figure were all annealed at $550^{\circ}\mathrm{C}$ for 30~s.
\textcolor{change}{(a)-(c) Voltage vs.\ current measurements of 3 nominally identical devices. }%, measured \textcolor{change}{at $T=4.2$~K}.
(d)-(f), Same data as in (a)-(c) plotted on a log-log scale.
For each dataset the critical current, $\mathrm{I_{C}}$, is marked.
Values where a negative voltage, below the noise floor \textcolor{change}{($\vert V\vert <1~\mu\mathrm{V}$)}, was measured are colored red. 
\textcolor{change}{The resistance}, indicated by the solid black line, is determined as the coefficient of a linear form of the data below $\mathrm{I_{C}}/2$.
\textcolor{change}{For (f) (the sample with the lowest resistance) the resistance is less than $R=V/I (I_{C})=0.35~\mathrm{m\Omega}$.}
\textcolor{change}{(g) The resistance at $T=4.2$~K vs. $\mathrm{I_{C}}$, acquired by repeating the above procedure for 21 nominally identical devices.} 
The results for (d), (e) \& (f) are identified by a red diamond, a green circle and a yellow square respectively. 
\textcolor{change}{The resistances of these nominally identical devices} varies by more than 5 orders of magnitude, while the critical current varies by an order of magnitude. 
This variation is even shown in devices fabricated on \textcolor{change}{the same die}. 
}
\end{figure}

Figures~\ref{fig:variation}(a)-(c) show sourced current vs.\ measured voltage for 3 nominally identical devices annealed at $550^{\circ}\mathrm{C}$ for 30~s and measured in liquid helium (at 4.2~K), on a lin-lin plot and figures~\ref{fig:variation}(d)-(f) show the same data, on a log-log plot.
Each plot has the same range and dimensions.
For each measurement the critical current, $\mathrm{I_C}$, at which the voltage increase becomes superlinear with increased current, is marked.
%For each measurement a current, $\mathrm{I_{C}}$, at which the voltage increase becomes superlinear with increased current, is marked.
%For high values of $\mathrm{I^*}$, $\mathrm{I^*}$ matches the critical current $\mathrm{I_C}$.
%A black line, which is a linear form of the voltage measured at currents below half $\mathrm{I_C}$. We call the coefficient of this linear form the residual resistance. 
A black line, corresponding to a linear relationship of the voltage measured to the current for currents below half $\mathrm{I_{C}}$, is also shown, and we call this slope the resistance at $T=4.2$~K. 
%At currents $>10~\mathrm{\mu A}$ the voltage change becomes superlinear with current,
For samples with a high resistance at $T=4.2$~K the voltage change eventually becomes linear again. 
For devices with low resistance at $T=4.2$~K this superlinear voltage change with current has not stopped within the measurement range of the experiment. 
Figure~\ref{fig:variation}(g) shows the resistance at $T=4.2$~K vs. $\mathrm{I_C}$ for 21 nominally identical devices. The resistance at $T=4.2$~K of these devices varies by more than 5 orders of magnitude while the critical current varies by an order of magnitude, even though these devices are nominally identical, some even located on the same $5\times5~\mathrm{mm}^2$ die.

We will now argue why the current at which the voltage behaviour becomes superlinear with current can be called the critical current even though the resistance of the devices is not zero below this current value, by modeling our system as a network of superconductors and normal resistors.
Let us limit our discussion to a simple network, an ensemble of superconductors and resistors connected in series. We denote the resistance of each normal resistor $r_{nn}$, the total normal resistance ${R_{nn}=\sum{r_{nn}}}$, the normal state resistance of each superconductor $r_{sn}$, and the total normal state resistance of the superconductors ${R_{sn}=\sum{r_{sn}}}$. 
If all the superconductors have the same critical current then we would expect the resistance of the network to change from $R_{nn}$ to ${R_{nn}+R_{sn}}$ when the critical current is exceeded. 
However, the precipitates vary in size, so we expect their critical currents to also vary.
Then instead of a sharp transition at a single current value from $R_{nn}$ to ${R_{nn}+R_{sn}}$ we would expect the resistance to be $R_{nn}$ below the lowest critical current and then start increasing as we go though the different critical current values of the precipitates, eventually reaching $R_{nn}+R_{sn}$ as we exceed the highest critical current value.
According to this model, what we have called the critical current in figure~\ref{fig:variation} is the lowest critical current of superconducting gallium nano-precipitates in the network.

%From figure~\ref{fig:variation}(g) there seems to be a correlation between higher resistance at $T=4.2$~K and lower critical currents. A possible explanation for this correlation is that for a given current value, devices with a higher resistance at $T=4.2$~K generate more heat than devices with a lower resistance at $T=4.2$~K. This current generated heat may then increase the temperature of the device, pushing it out of the superconducting regime, at lower currents for devices with a higher resistance at $T=4.2$~K. 

\section{Final oxide thicknesses} % (fold)
\label{sec:final_oxide_thickness}

As discussed in section~\ref{sec:sample_prep} all devices had nominally 30~nm of $\mathrm{SiO}_2$ sputtered onto them before the Ga ion implantation and during the ion implantation a portion of this oxide is ablated. 
Furthermore, different amounts of additional gate oxide were added during subsequent processing for different devices. 
Determining the final oxide thickness is important in order to accurately determine the applied external electric field from the applied top gate voltage.
For devices where the additional gate oxide was comparable to the initial 30~nm of $\mathrm{SiO}_2$ sputtered prior to the Ga ion implantation, scanning transmission electron microscope (STEM) images were used to determine the final oxide thickness (the preparation and imaging of STEM samples is discussed section~\ref{sub:stem_sample_preparation}). From these STEM images we can determine that most of the sputtered oxide is ablated during the ion implantation (see figure~\ref{fig:methods}(e)).
For devices where the additional gate oxide was thicker ($\geq100~\mathrm{nm}$) than the initial 30~nm of $\mathrm{SiO}_2$ sputtered oxide we use the result that most of the sputtered oxide is ablated during the ion implantation, and assume that the final oxide thickness can be approximated, within $\sim 10\%$, by the thickness of the additional oxide.
Finally, we note that as we only observe negligible changes in device resistances with applied external electric fields, accurately determining the value of the external field is not critical to the results shown nor the conclusions drawn in this paper.

% section final_oxide_thickness (end)

%\section*{References}

%\bibliographystyle{iopart-num}

\end{document}